%% file: main.tex
\documentclass[conference]{IEEEtran}
\IEEEoverridecommandlockouts

\usepackage{cite}
\usepackage{amsmath,amssymb,amsfonts}
\usepackage{algorithmic}
\usepackage{graphicx}
\usepackage{textcomp}
\usepackage{comment}
\usepackage{xcolor}
\usepackage{booktabs}
\usepackage[hyphens]{url}
\usepackage{multirow}
\def\BibTeX{{\rm B\kern-.05em{\sc i\kern-.025em b}\kern-.08em
    T\kern-.1667em\lower.7ex\hbox{E}\kern-.125emX}}
    
\begin{document}

\title{Political Narratives \\in the 2024 Moldovan Presidential Election: \\Stance, Amplification, and Influence
}

\author{\IEEEauthorblockN{1\textsuperscript{st} Andrea Piras}
\IEEEauthorblockA{\textit{Department of Computer Science} \\
\textit{University of Illinois Chicago}\\
Chicago, USA \\
apiras2@uic.edu}
\and
\IEEEauthorblockN{2\textsuperscript{nd} Osman Rahmoun}
\IEEEauthorblockA{\textit{Department of Computer Science} \\
\textit{University of Illinois Chicago}\\
Chicago, USA \\
orahm2@uic.edu}
\and
\IEEEauthorblockN{3\textsuperscript{rd} Elena Zheleva}
\IEEEauthorblockA{\textit{Department of Computer Science} \\
\textit{University of Illinois Chicago}\\
Chicago, USA \\
ezheleva@uic.edu}
}

\makeatletter
\def\@IEEEpubidpullup{4\baselineskip}
\makeatother

\IEEEpubid{%
\makebox[\textwidth][l]{%
    \parbox[b]{\columnwidth}{%
        \scriptsize
        \raggedright
        \copyright~2026 IEEE. Personal use of this material is permitted.
        Permission from IEEE must be obtained for all other uses, in any
        current or future media, including reprinting/republishing this
        material for advertising or promotional purposes, creating new
        collective works, for resale or redistribution to servers or lists,
        or reuse of any copyrighted component of this work in other works.
    }%
}}

\maketitle


\begin{abstract}
Social media platforms play a central role in shaping political discourse during elections, particularly in geopolitically contested regions. We study how political narratives emerged, aligned, and propagated during the 2024 Moldovan presidential election by analyzing publicly available posts collected from X, Facebook, and Reddit between August and October 2024. Combining multilingual stance detection, coordination analysis, and causal inference, we examine both descriptive and causal mechanisms underlying online political discourse.
First, we apply large language models to classify stance toward European integration, Russia, and the two main presidential candidates, revealing clear cross-platform and cross-linguistic differences in how political positions are expressed. Second, we construct retweet-based and semantic alignment networks on X to identify amplification and aligned framing of dominating narratives. Finally, we use a proximal causal inference framework to estimate peer influence effects in stance formation.
Our results indicate consistently positive and statistically significant average treatment effects across major stance categories. These findings provide evidence that amplification and peer exposure contribute to the reinforcement of political narratives online. This work offers the first analysis of social media discourse surrounding the 2024 Moldovan presidential election and demonstrates the value of integrating language-based annotation, network analysis, and causal inference for studying political influence in multilingual and geopolitically sensitive contexts.
\end{abstract}


\input{sections/1-introduction}

\input{sections/2-related_work}
\input{sections/3-data}
\input{sections/4-methods_stance}
\input{sections/5-methods_coordination}
\input{sections/6-methods_causal}

\input{sections/7-discussion}

\bibliographystyle{IEEEtran}
\bibliography{IEEEabrv,references}

\clearpage
\input{sections/appendix}

\end{document}

%% file: sections/1-introduction.tex
\section{Introduction}

Social media platforms have become pivotal places for political contestation, shaping both the circulation of information and the formation of public opinion during elections. Research has shown that online discussions mirror and amplify offline divisions, with coordinated campaigns and partisan hostility emerging as key features of digital political debate~\cite{barbera2015birds, ccetinkaya2025cross}. At the same time, platforms serve as spaces where diverse actors, including citizens, journalists, and anonymous accounts, compete to push narratives and provide support. Understanding these dynamics is fundamental in elections held in strategically contested regions, where geopolitical orientations can influence outcomes that extend beyond the national borders.

The 2024 Moldovan presidential election is a clear example of this dynamic. Positioned between Europe and Russia, Moldova has long been at the center of political and cultural controversies. This election became a focal point for debates over European integration and Russian influence, particularly between the two main Moldovan candidates, Maia Sandu (who ultimately won the election) and former Prosecutor General Alexandr Stoianoglo.
Sandu is often described by major outlets as pro-European, as her campaign was tied to EU integration, including the EU accession referendum held alongside the first round. Stoianoglo, on the other hand, is widely described as backed by a party traditionally linked to Russia (PSRM), while publicly expressing opposition against European integration. 

In this paper, we present an analysis of election-related social media posts collected during the election period (between August~1 and October~25,~2024) from three social media sites: X (formerly Twitter), Facebook, and Reddit. To the best of our knowledge, no prior work has conducted a comprehensive and systematic analysis of social media discourse surrounding the 2024 Moldovan presidential election. Our approach integrates stance detection, network analysis, and amplification measures to examine how narratives and alignments unfolded online. To identify both consistent and contradictory cases that reveal the fluidity of political discourse, we perform several analyses. First, we employ large language models (LLMs) to classify stance toward Europe, Russia, and the Moldovan candidates Sandu and Stoianoglo. Second, we construct temporally proximate retweet and semantic similarity networks, which capture amplification and content alignment across communities. These networks allow us to identify central actors, information flows, and potential signals of aligned narrative activity. Finally, we employ causal analysis to understand which stances are influential and contribute to social contagion.

Specifically, in this paper, we address the following questions related to political discussion on social media:
\begin{itemize}
    \item \textbf{Distribution of political stances (RQ1)}: How were stances aligned with the main candidates’ positions on EU integration and relations with Russia distributed across election-related online discourse?
    \item \textbf{Amplification dynamics (RQ2)}: To what extent do amplification signals, e.g., rapid retweets and temporally proximate sharing of similar tweets, shape narrative visibility and alignment during the elections?
    \item \textbf{Network influence of stances (RQ3)}: Does exposure to tweets with a political stance (e.g., pro-Europe, anti-Russia) causally increase the probability that a user later expresses the same stance?
\end{itemize}

Our contributions are: (1) we provide the first systematic account of social media discourse around the 2024 Moldovan presidential elections, (2) we combine stance classification with amplification analyses to methodologically analyze electoral data, and (3) we evaluate model-based estimates of peer-exposure effects to understand the spread of ideas in the network.

Our code, LLM prompts, and tools are available in the public repository: \url{https://github.com/edgeslab/moldova-elections}.

%% file: sections/2-related_work.tex
\section{Related Work}
Research at the intersection of political communication and computational social science has extensively examined how social media platforms shape, mediate, and amplify political ideas during elections~\cite{conover2011political, stier2020election}. Platforms like X serve not only as channels of information but as spaces where politicians, media, and ordinary users compete to shape the conversation. 
In computational approaches, such dynamics are often measured through network structure and content signals. For example,~\cite{jiang2023retweet} propose Retweet-BERT, a model that infers political leaning from language and retweet patterns, showing how textual and structural features can jointly reveal ideological alignments. Different studies have documented polarization across major electoral events, such as U.S. presidential campaigns~\cite{conover2011political,barbera2015birds} and European elections~\cite{stier2025online, muller2022real, nizzoli2021coordinated}, illustrating how political cleavages are mirrored in network connectivity, stance expression, and engagement. Similar work examines the effect of political discourse on echo chambers, self-reinforcing network structures in which users are disproportionately exposed to ideologically aligned content~\cite{garimella2018political, cinelli2021echo, cinus2022effect}. Other work on polarization highlights how emotionally-charged content drives engagement between groups~\cite{brady2017emotion, gao2024crisis}, shaping the diffusion of political narratives online.

A second group of research focuses on coordinated behavior and amplification, a central theme in recent studies of campaign dynamics. Synchronization in posting activity, especially through retweets and news-sharing, can signal organized efforts to boost specific narratives or actors~\cite{burghardt2024socio, zhu2024study}. Classic approaches leveraged engagement networks to classify coordinated campaigns~\cite{pacheco2020uncovering, pacheco2020whitehelmets}. More recent studies identified coordination by detecting temporal clustering, co-retweeting, and domain-sharing spikes~\cite{nizzoli2021coordinated}. 

A third group of work focuses on stance detection and topic modeling, both of which are widely used to map user attitudes and ideological alignments. Stance detection has evolved from lexicon-based and supervised approaches~\cite{augenstein2016usfd}, to zero-shot transformer-based methods capable of scaling across domains~\cite{taranukhin2024stance}. This shift has enabled researchers to analyze topics in low-resource contexts where labeled data are scarce. Cross-lingual stance and topic detection techniques~\cite{zhang2023target, bianchi2021cross} have enabled the examination of how narratives are framed across linguistic communities. Applications to politics are common, with work analyzing major political elections~\cite{muller2022real} and events, such as international conflicts~\cite{kloo2024cross}. 

Finally, a growing line of research applies causal inference methods to social platforms to estimate how exposure to information shapes user behavior. Traditional observational analyses are often limited by latent homophily, making it difficult to distinguish peer influence from shared preferences. Recent work has used matching and instrumental variable approaches to separate influence from selection in network data~\cite{aral2009distinguishing}, often in the context of social contagion and diffusion. 
Similar work demonstrates that high-dimensional adjustment can reduce bias relative to experimental benchmarks~\cite{shalizi2011homophily,eckles2017design}.
More recently, attention has shifted toward the role of interference in networked settings.~\cite{fatemi2024cascade} show that cluster-based randomization can be biased in settings where influence propagates through multi-hop cascades. They propose a cascade-aware experimental design that underscores the importance of accounting for this interference in studies of social media influence.
Recent methodological advances also focus on estimating individual and heterogeneous causal effects under complex peer influence mechanisms, including approaches that explicitly model the influence of neighbors’ behaviors on outcomes~\cite{adhikari2025inferring, adhikari2025learning}, but do not address hidden confounding.
Within this context,~\cite{fatemi2025contagion} combine representation learning and balancing techniques to estimate contagion effects under hidden confounding.

Our work brings these works together by combining multilingual topic and political stance detection, coordination analysis, and causal effect estimation in the context of a politically sensitive election that has received little prior attention. Unlike other studies, which focused on major Western electoral contexts, this work examines an underexplored region with complex linguistic and geopolitical divisions, offering new insights into how narratives spread, align, and gain traction during political campaigns.

%% file: sections/3-data.tex
\section{Data}

Our study analyzes election-related discourse surrounding the 2024 Moldovan presidential election using social media data obtained as part of a broader, grant-funded data collection program. The program used a third-party provider to collect election-related posts from X, Facebook, and Reddit based on specific keywords (provided in the Appendix). The data was accessed in accordance with platform terms of service and regulatory constraints. 
The program did not reveal the budgetary constraints of the data collection or the name of the data provider. The extraction strategy was designed by the program to maximize topical and actor-level coverage of election-related discourse across platforms. As a result, the completeness of interaction data vary by platform: X provides the most comprehensive interaction data (retweets), whereas Reddit and Facebook offer limited engagement information beyond each individual post. The dataset consists of 52,696 publicly available posts posted between August~1 and October~25,~2024, covering the main campaign period and the first round of the Moldovan presidential election. Daily volume remains relatively low from August through mid-October, followed by a sharp spike around the election date on October~20,~2024. Approximately 34.7\% of all collected posts were published within the three-day election window from October 19-21, indicating that a substantial share of the dataset captures discourse during the immediate election-period surge.


After filtering duplicate content and posts without textual content, X includes 45,695 posts from 30,940 unique accounts. When available, missing parent tweets were retrieved to preserve conversational context and enable analyses of retweet and reply behavior. Each tweet is associated with its full text, timestamp, detected language, and engagement metrics, including retweets, quotes, replies, and likes.
Reddit and Facebook data provide complementary perspectives, albeit with less information available on engagement and interactions (e.g., reposts). Reddit includes 3,555 posts and comments from 2,812 unique accounts, which contain textual content, timestamps, and author identifiers, allowing partial reconstruction of reply relationships, though parent–child links are incomplete for a substantial fraction of comments. Facebook includes 3,446 posts from 393 users, containing post text and timestamps, but lacking information on reshares, replies, or interaction networks. As a result, network and causal analyses requiring fine-grained interaction structure are restricted to X.

\begin{figure}[ht]
\centering
\includegraphics[width=1\columnwidth]{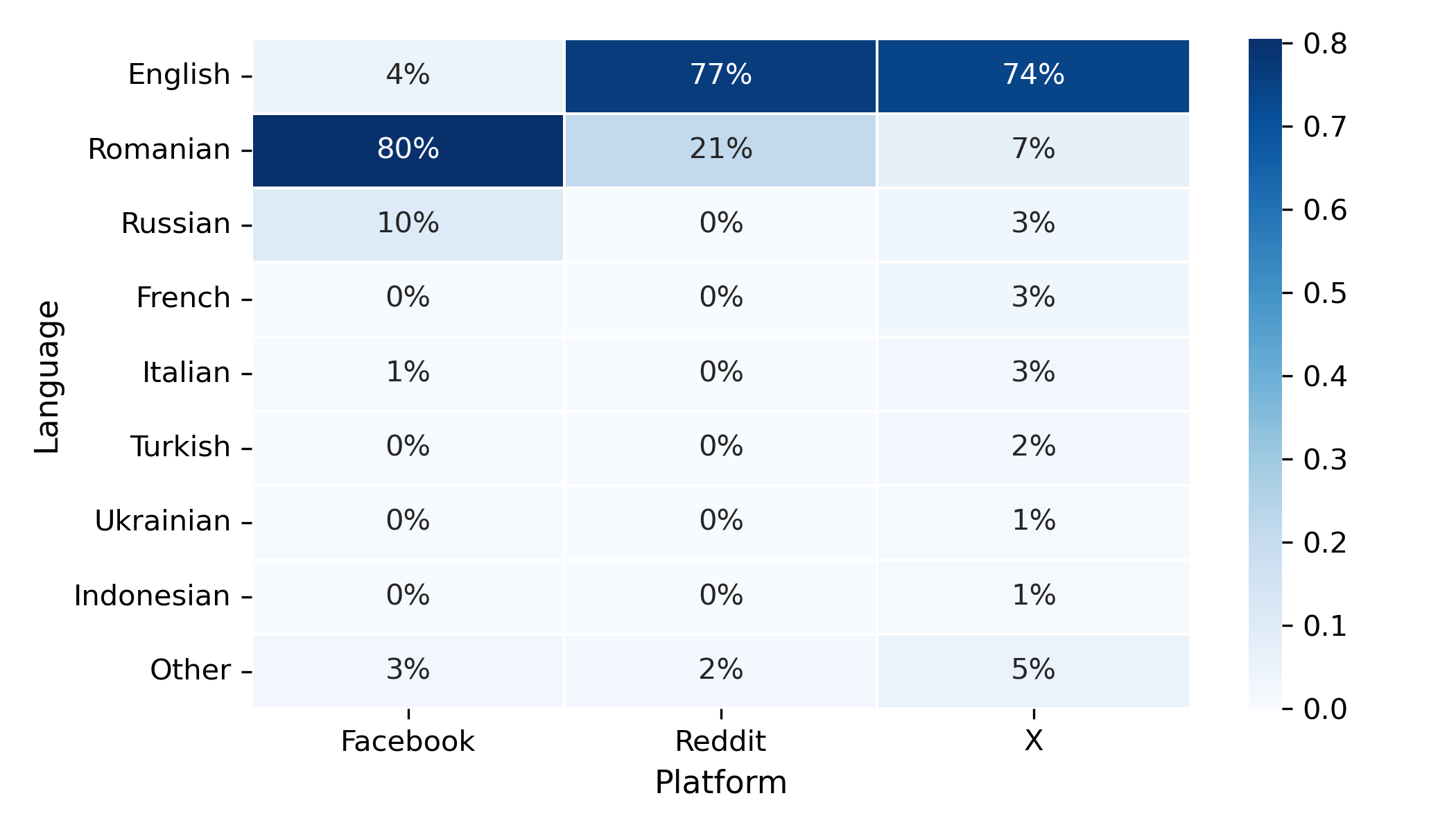}
\caption{Normalized heatmap of language share within each platform. Only the eight most frequent languages are shown individually; all remaining languages are aggregated into an ``Other'' category. For Facebook, 80\% of posts are in Romanian, while for Reddit and X, more than 75\% of posts are in English.}
\label{fig:languages}
\end{figure}

Fig.~\ref{fig:languages} reports the distribution of languages across platforms, highlighting the prominence of Romanian (80\%) and Russian (10\%) in Facebook posts, and of English (77\% and 75\%) and Romanian (21\% and 7\%) within Reddit and X communities.

Text preprocessing included lowercasing, punctuation removal, language detection, and machine translation for non-English content, enabling cross-lingual analysis. Multilingual sentence embeddings were generated for all posts using the \texttt{SentenceTransformer-paraphrase multilingual-mpnet-base-v2} model~\cite{reimers-emnlp19}.

To uncover the themes driving political debates, we first performed topic modeling on the posts from all three platforms. 
We adopted an embedding-based topic extraction pipeline as implemented in BERTopic~\cite{grootendorst-arxiv22}, which has been shown to perform better than other topic models~\cite{egger-sociology22}. This approach provides fine-grained yet interpretable clusters and is more robust to the linguistic variability and fragmentary nature of social media posts, making it well-suited to the structure of our dataset. While BERTopic produces high-quality clusters and representative keywords, it does not automatically generate human-readable topic labels. To do so, we extracted the most informative keywords for each cluster based on their TF–IDF scores and used these as the basis for expert interpretation. Domain experts reviewed the top-ranked words and associated post samples for each cluster to assign short, human-readable labels summarizing the main theme. Finally, we excluded topics deemed not relevant to the elections.
The top three identified topics (``Moldova, EU membership, Russia, and regional identity", ``Ukraine, Russia, NATO, and post-Soviet geopolitics", and ``President Maia Sandu and Moldovan presidential elections") account for $64\%$ of all posts in the dataset. The complete list of topics and their frequencies can be found in the Appendix.



Fig.~\ref{fig:topics_platform} further compares the relative prevalence of major topics across platforms, restricting attention to topics that account for at least $2\%$ of posts on at least one platform. 

\begin{figure}[ht]
\centering
\includegraphics[width=1\columnwidth]{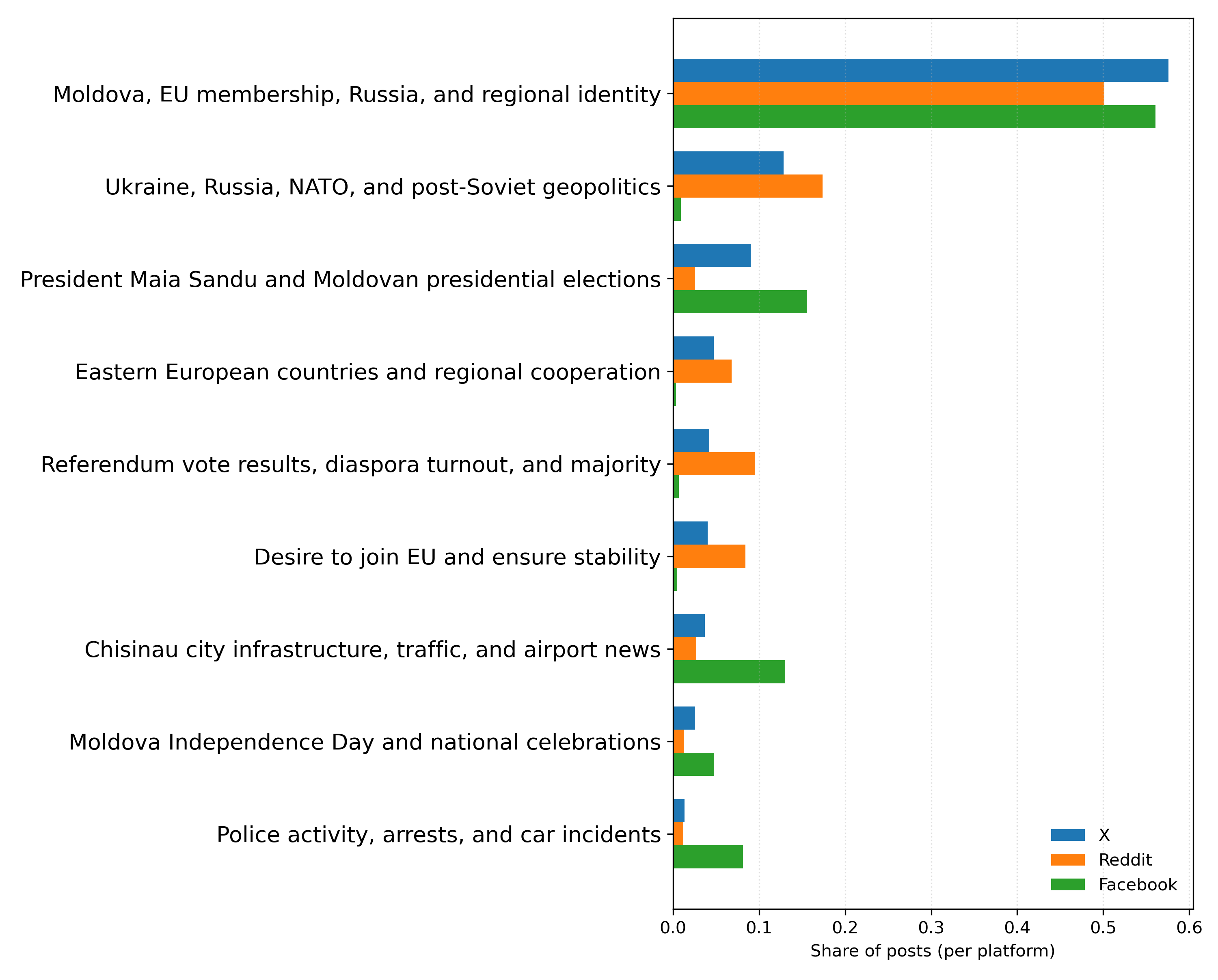}
\caption{Distribution of topics accounting for at least $2\%$ of posts on at least one platform. X and Reddit communities display similar attention to geopolitical issues, whereas Facebook users predominantly engage with discussions centered on domestic governance.
}
\label{fig:topics_platform}
\end{figure}

\begin{figure*}[h!]
\centering
\includegraphics[width=0.9\linewidth]{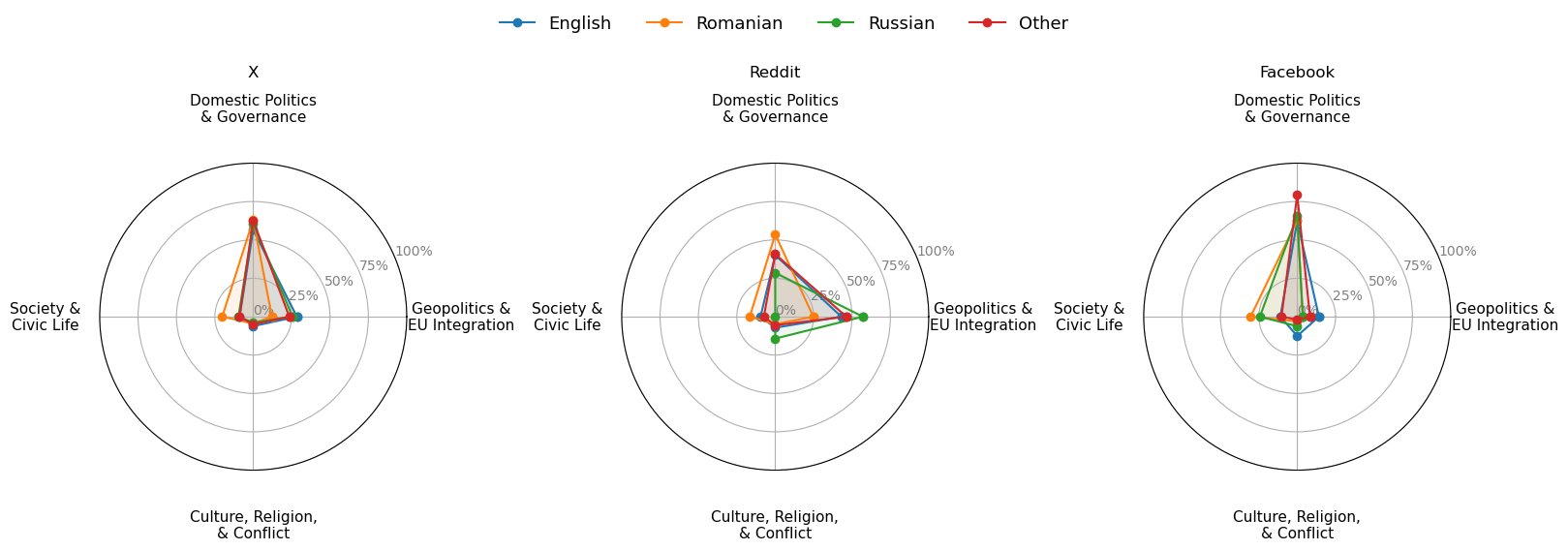}
\caption{Distribution of macro-topics within each language community across platforms. Across X, Facebook, and Reddit, ``Domestic Politics \& Governance" accounts for the largest share of posts in all language groups, particularly on Facebook, where it represents $60\%$-$80\%$ of content. ``Geopolitics \& EU Integration" is comparatively more prominent among English and Russian posts, especially on Reddit and X. ``Society \& Civic Life" receives greater attention within Romanian-language discourse across platforms, while ``Culture, Religion, \& Conflict" consistently represents a small fraction of discussion.}
\label{fig:topics_language}
\end{figure*}

Across all three platforms, discussion is strongly dominated by a small set of high-level themes, most notably the one related to Moldova’s geopolitical positioning, European integration, and relations with Russia. This topic consistently represents the single largest share of posts on X, Reddit, and Facebook ($57\%$, $55\%$, and $50\%$ of posts, respectively), indicating a shared core focus across platforms. Beyond this dominant theme, however, the distribution of attention varies by platform. On X and Reddit, secondary topics remain relatively broad and internationally oriented, with substantial attention devoted to Ukraine-Russia geopolitics and NATO, as well as to broader regional dynamics. In contrast, Facebook shows greater emphasis on domestic political processes, including referendum outcomes, voter turnout, Chisinau events, and campaign-specific developments. 
Fig.~\ref{fig:topics_language} further analyzes these differences by showing how language communities are distributed across topics within each platform. For interpretability, we manually grouped fine-grained topics into a smaller set of macro-topics capturing broader narrative categories: 1) ``Geopolitics and EU Integration"; 2) ``Society and Civil Life"; 3) ``Domestic Politics and Governance"; 4) ``Culture, Religion, and Conflict". The topic grouping and mapping procedure is in the Appendix.
On X, the majority of posts for each language ($>60\%$) focus on ``Domestic Politics and Governance", indicating that election-related discourse is primarily framed around internal political dynamics across language communities. The Romanian community then focuses on ``Society and Civil Life" ($\sim20\%$), suggesting greater attention to the election's social and civic implications.
A similar but more balanced pattern appears on Reddit for the Romanian, English, and ``Other" language communities ($25-50\%$ for ``Domestic Politics and Governance" and ``Geopolitics and EU Integration"). In contrast, Russian-language posts focus primarily on ``Geopolitics and EU Integration" ($\sim60\%$), consistent with a discourse that accentuates external relations over internal governance. 
On Facebook, all language communities show a stronger concentration on ``Domestic Politics and Governance" ($60-80\%$ of the share), reinforcing the platform’s role as a space for locally oriented political discussion, with a secondary share ($\sim25\%$) of ``Society and Civil Life" posts among the Romanian and Russian communities.
Across all languages and platforms, ``Culture, Religion, \& Conflict" consistently represents the least-discussed topic, accounting for only a small fraction within each language community.

%% file: sections/4-methods_stance.tex
\section{Distribution of political stances (RQ1)}

A central goal of this work is to characterize the emergence and evolution of geopolitical alignments during the 2024 Moldovan presidential election. To this end, we identify political stance expressed at the tweet level and analyze how these alignments vary over time and across platforms and language communities.
Each tweet was annotated along two primary axes reflecting the main political cleavage in the Moldovan election: stance toward the European Union and/or the presidential candidate Maia Sandu (which we call the ``Europe" stance), and stance toward Russia and/or the candidate Alexandr Stoianoglo (which we call the ``Russia" stance). For each axis, tweets were assigned one of three labels: \emph{pro}, \emph{anti}, or \emph{neutral}. Posts were allowed to receive multiple labels across axes (e.g., ``pro-Europe" and ``anti-Russia"), reflecting the mixed framing of political discourse during the campaign.
Political stance annotation was performed using a large language model (LLM), Gemini~\cite{comanici2025gemini}, through a structured prompting pipeline. We employed a few-shot prompting strategy augmented with chain-of-thought (CoT) reasoning~\cite{sahoo2024systematic}, in which the model was provided with concise labeled examples and instructed to produce both a short rationale and a final stance label for each target dimension. This design was chosen to improve consistency and robustness in the presence of sarcasm and indirect political references, which are common in election-related social media content.
We evaluated multiple prompting configurations, including zero-shot, few-shot without reasoning, and few-shot with CoT, against a manually annotated validation set of 200 randomly selected tweets curated by two domain experts, with disagreements adjudicated through a consensus process. Accuracy was computed as the proportion of model predictions whose final stance label matched the annotated human label. The CoT few-shot configuration achieved the highest agreement and stable accuracy exceeding 90\%, and was therefore applied to the full dataset. More information on the labeling process and the different prompting configurations can be found in the Appendix. Ambiguous or seemingly contradictory cases (e.g., simultaneous pro- and anti-stance assignments) were retained rather than collapsed, as manual inspection indicated that such cases often arise from quoting, irony, or mixed framing.
The resulting stance annotations form the basis for all downstream analyses, including temporal trend analysis, coordination and network-based studies, and causal effect estimation.

\begin{figure}[ht]
    \centering
    \includegraphics[width=\linewidth]{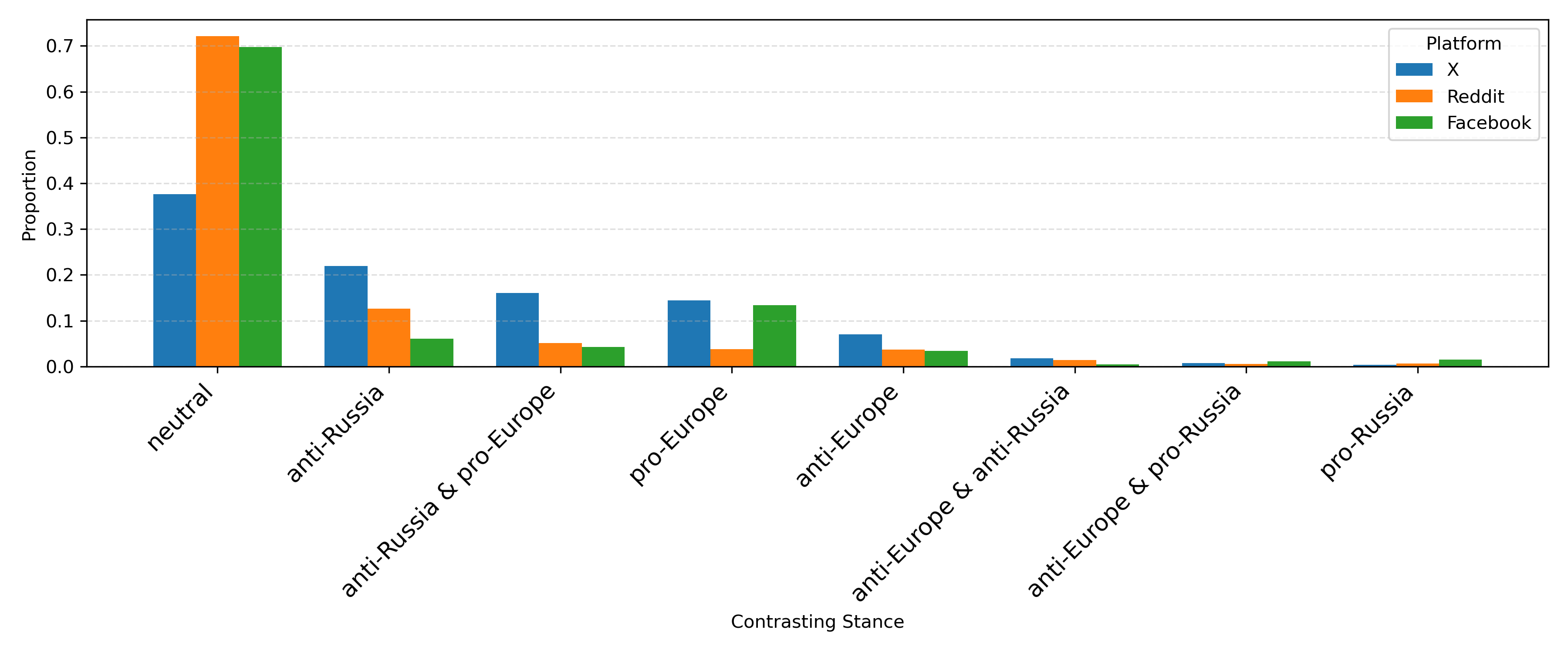}
    \caption{Aggregate distribution of major political stances expressed in election-related posts. Excluding neutral stance, ``anti-Russia", ``anti-Russia \& pro-Europe", and ``pro-Europe" stances constitute the majority of observed content ($5-20\%$ across all platforms), while ``pro-Russia" and ``anti-Europe" stances appear less frequently ($<5\%$).}
    \label{fig:stance_distribution}
\end{figure}

We first examine how contrasting geopolitical and candidate-aligned stances were distributed across election-related discourse. Fig.~\ref{fig:stance_distribution} reports the aggregate prevalence of each major stance category over the full observation period. 
Neutral posts represent the largest share of content ($43\%$ for X, and $\sim70\%$ for Reddit and Facebook). For non-neutral posts, the conversation is consistently dominated by ``anti-Russia", ``anti-Russia \& pro-Europe", and ``pro-Europe" framing, though the relative proportion differs by platform. X content has $20\%$ of ``anti-Russia" posts, and similarly $13\%$ of both ``anti-Russia \& pro-Europe" and ``pro-Europe" posts, suggesting a discourse that frequently leans towards favoring a European stance. Reddit shows a less balanced profile, with ``anti-Russia" occupying $13\%$ of content, while the remaining stances each account for less than $5\%$ of posts. Facebook, in contrast, places greater weight on ``pro-Europe" ($13\%$) than on ``anti-Russia" and ``anti-Russia \& pro-Europe", each amounting to less than $5\%$ of posts. Explicitly ``pro-Russia" or ``anti-Europe" positions remain rare across platforms.
\begin{figure}[ht]
    \centering
    \includegraphics[width=\linewidth]{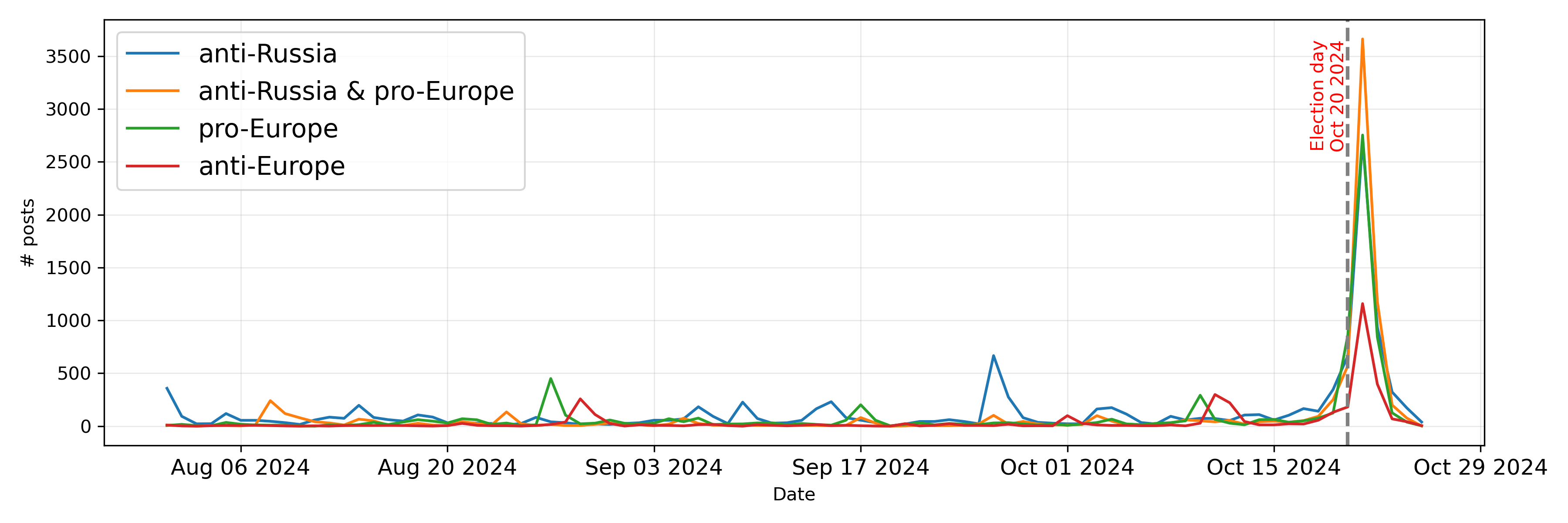}
    \caption{Daily volume of stance-labeled posts over time. The vertical line marks the first round of the Moldovan presidential election on October~20,~2024. ``pro-Europe", ``anti-Russia", and ``anti-Russia \& pro-Europe" stances exhibit the most pronounced post-election surge.}
    \label{fig:stance_time}
\end{figure}

To examine temporal dynamics, we focus on non-neutral stance categories that account for at least $5\%$ of posts in at least one of the social media, ensuring sufficient volume for reliable trend analysis. Fig.~\ref{fig:stance_time} shows the daily volume of these dominant stances over time. Two patterns stand out. First, posts with ``pro-Europe" and ``anti-Russia \& pro-Europe" stances show pronounced temporal variability, with increases to 3,600 and 2,700 posts shared immediately after the first round of the election on October~20,~2024. Second, the composition of this peak is highly asymmetric: posts expressing ``pro-Europe", ``anti-Russia", and ``anti-Russia \& pro-Europe" stances account for the vast majority of the surge, with activity increasing by orders of magnitude relative to earlier baselines. In contrast, ``anti-Europe" content also rises but remains substantially lower in volume, peaking at approximately 1,200 posts.

%% file: sections/5-methods_coordination.tex
\section{Amplification Dynamics (RQ2)}
To examine how narratives were amplified and aligned during the 2024 Moldovan presidential elections, we conducted a amplification analysis on X accounts. Given the lack of reshare information on Reddit and Facebook, we excluded these platforms from the analysis. Amplification can emerge both organically (e.g., reactions to a breaking event) and strategically (e.g., organized campaigns or information operations). Identifying these patterns provides insight into how information is propagated collectively, which actors contribute to narrative diffusion, and how different groups structure their activity for visibility.
Our analysis builds on two complementary amplification signals, temporal synchronization in retweeting and message posting, which reflect behavior in amplifying or repeating messages.
Inspired by the methodology proposed by~\cite{de2024detecting} and~\cite{pacheco2020uncovering}, we analyze these signals using rapid-retweet and rapid semantic-similarity networks.


\noindent\textbf{Rapid Retweet Networks.} We first constructed rapid retweet networks. 
To select the temporal threshold, we conducted a sensitivity analysis following the approach of~\cite{de2024detecting}. We evaluated three candidate values for $\Delta t$ (60\,s, 1\,hour, 24\,hours) and examined their effects on network density, clustering coefficient, and edge weight distribution (Analysis shown in the Appendix). The 60\,s threshold provides the best signal for rapid synchronization and coordination (similar works use up to 5\,minutes as the threshold~\cite{ng2022online, magelinski2022synchronized}), but covers only a small fraction of the retweet activity and yields a much smaller network. We therefore use 1\,hour as the primary descriptive threshold for event-scale synchronized engagement, while interpreting it as a broad amplification signal rather than evidence of orchestrated coordination. 
\begin{figure}[ht]
    \centering
    \includegraphics[width=0.85\linewidth]{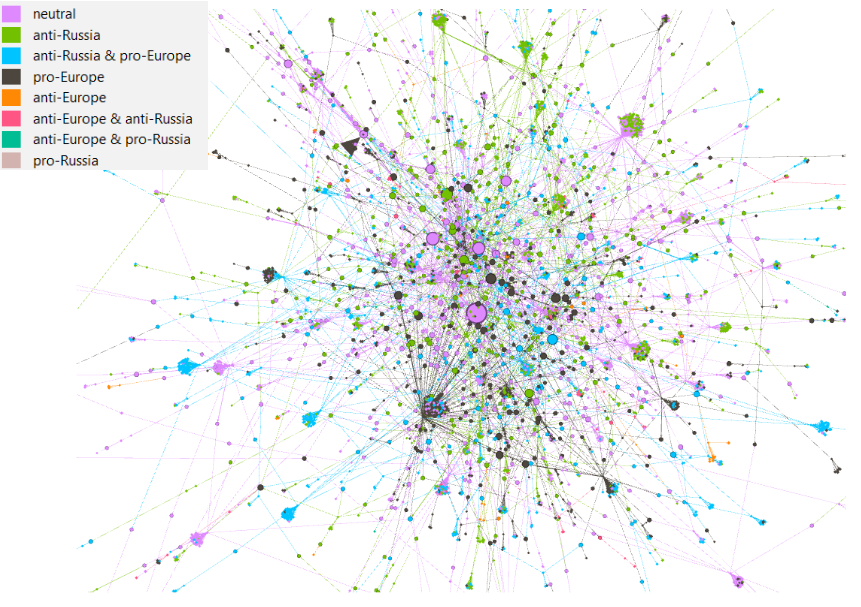}
    \caption{Rapid retweet subgraph using a 1-hour temporal threshold. Nodes represent users; edges connect users who retweeted another user's content within the threshold. The node size reflects the number of interacting users.}
    \label{fig:retweet_network}
\end{figure}
The resulting rapid retweet network, constructed using the 1-hour temporal threshold, exhibits a sparse global structure with a pronounced, more connected core of 5,501 nodes with an average degree of 2.32. Fig.~\ref{fig:retweet_network} shows the core subgraph with nodes representing users and edges indicating amplification, through retweets, of the same content within the 1-hour window. This central component concentrates a large fraction of amplification activity and is surrounded by smaller, weakly connected clusters (1-5 users). Manual inspection of the highest-degree nodes in the core reveals that they correspond primarily to prominent political actors, international news outlets, journalists, activist accounts, and satirical collectives. Examples include official political accounts (e.g., @SanduMaiaMd, @vonderleyen), Russo-Ukrainian war reporting accounts (e.g., @wartranslated), journalists monitoring Russian state media (e.g., @JuliaDavisNews), and opposition-aligned Eastern European news outlets (e.g., @visegrad24). 
Despite their heterogeneity in role and affiliation, these accounts overwhelmingly express or amplify ``anti-Russia" and ``pro-Europe" narratives. Illustrative examples are reported in the Appendix.
Thus, amplification during the election was not evenly distributed across stances: narratives from news outlet and political actors opposing Russian influence and supporting European integration were the primary objects of amplification. In contrast, ``pro-Russia" and ``anti-Europe" content appears in smaller, more fragmented clusters that remain weakly connected to the central component, suggesting lower visibility within this retweet-based structure.


\noindent\textbf{Rapid Semantic Networks.} We extended this logic to original tweet content by detecting temporally proximate posting of semantically similar messages. This step captures cases where accounts may not retweet the same post but instead produce similar framing within the same event-scale time window, highlighting semantic alignment patterns that can arise from coordinated messaging, shared attention to events, or common narratives. We first removed retweets from the dataset, as this would cause same-text posts to be connected in the resulting network, without providing any additional insight. We then generated multilingual sentence embeddings for all tweets and computed pairwise cosine similarity. Tweets exceeding a cosine similarity threshold and posted within the same temporal window were clustered, and accounts involved were connected in a semantic alignment network. While the retweet network captures synchronized amplification of the same content, the semantic network offers a complementary perspective on how similar narratives circulate across accounts.

We applied a cosine similarity threshold of 0.85, chosen through sensitivity analysis analogous to the retweet case. Lower thresholds (e.g., 0.70) resulted in overly dense graphs dominated by weak, noisy matches, while higher thresholds (e.g., 0.90) produced very sparse networks dominated by near-duplicate content. The 0.85 threshold balanced network interpretability and cluster detection, aligning with prior findings in coordinated behavior research~\cite{de2024detecting}. As with the retweet network, we interpret the 1-hour temporal window as capturing event-scale semantic alignment and trending information.

\begin{figure}[ht] 
\centering \includegraphics[width=\linewidth]{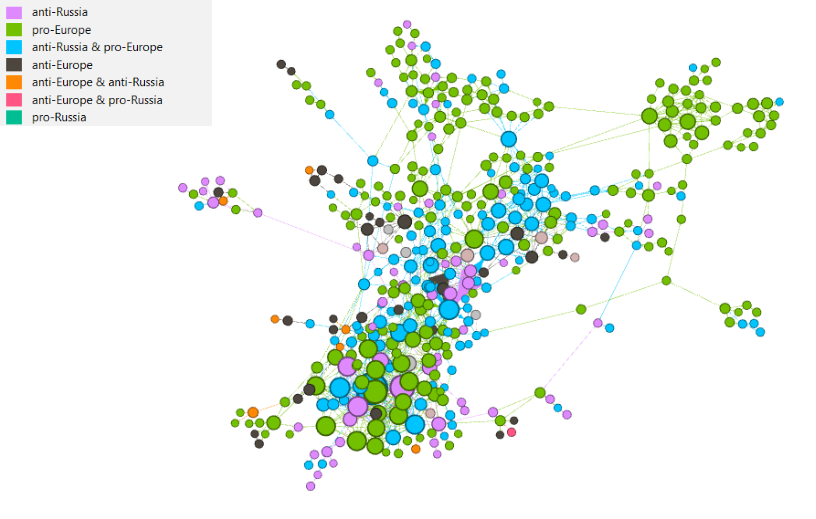} \caption{Semantic similarity alignment subgraph capturing posts that are both semantically similar and temporally proximate. Edges indicate high cosine similarity. The node size reflects the number of users posting similar tweets. The alignment clusters show a prevalence of ``pro-Europe" discourse.} \label{fig:semantic_network} 
\end{figure}

Compared to the rapid retweet network, this graph is substantially smaller (436 users) and forms a single, more interconnected core, surrounded by a larger set of small clusters (1-2 users), albeit with a higher average degree of 5.5. This indicates that semantic alignment captures a narrower set of behaviors, primarily reflecting near-duplicate or closely related framing. Within this semantic network, we focused on alignment around non-neutral stances.
Fig.~\ref{fig:semantic_network} presents the rapid semantic similarity alignment network core, excluding the ``neutral" stance-aligned posts. Due to the lack of direct information about usernames, we were unable to infer knowledge about hub accounts at the center of the clusters. 

There is a concentration, particularly of ``pro-Europe" and, secondly, of ``anti-Russia \& pro-Europe" and ``anti-Russia" positions. The cluster corresponds to bursts of similar messaging around topics opposing Russian influence, including foreign interference, electoral legitimacy, and regional security.

%% file: sections/6-methods_causal.tex
\section{Causal Effects of Exposure on Stance Expression (RQ3)}
\label{sec:causal}

The final component of our analysis examines whether exposure to political stance content affects users’ subsequent expression of similar views. To move beyond correlation, we adopt a proximal causal inference framework~\cite{tchetgen2020introduction} that leverages high-dimensional proxies to estimate treatment effects under unobserved confounding. Specifically, we would like to understand whether certain stances are contagious and can influence people on social media. Unlike descriptive correlations between political language and engagement, measuring peer influence in social media requires accounting for latent confounders such as homophily and network clustering.

Our causal setup follows the proximal embeddings (ProEmb) framework for estimating contagion effects in social networks~\cite{fatemi2025contagion}. We focus on one stance dimension at a time (e.g., ``anti-Europe"), which we call the target stance, and construct tweet-level observations for each user. For each target stance, we define the \emph{outcome} for each tweet based on whether it expresses that stance (binary indicator based on our stance labels). The \emph{treatment} for that tweet captures whether the user had prior observable exposure to that stance through the user’s peers. We define peers as the authors of tweets that a user has retweeted in the past, and then aggregate all of these peers’ tweets posted before each ego tweet timestamp. Treatment is set to one if at least one of these peer tweets expresses the target stance; otherwise, it is zero. Our treatment estimation captures the effect of prior stance-aligned content produced by accounts to whom the ego has revealed attention through retweeting. As platform exposure logs are not available, this treatment definition should be interpreted as interacted exposure, rather than as complete exposure to all content a user may have seen.
For each target stance $s$, we define causal effects at the tweet level. Let $Y_{i,t}^{(s)}$ denote the outcome for an ego tweet posted by user $i$ at time $t$, representing whether the tweet expresses stance $s$. Let $T_{i,t}^{(s)}$ denote the treatment indicator capturing prior exposure to stance $s$ through the user’s peers. The individual treatment effect (ITE) measures how exposure to stance $s$ causally changes the expected outcome for a specific ego tweet, by comparing the counterfactual scenarios in which the tweet is exposed, $T_{i,t}^{(s)} = 1$, versus not exposed, $T_{i,t}^{(s)} = 0$: 

\scalebox{0.9}{$
\mathrm{ITE}_{i,t}^{(s)}
= \mathbb{E}\!\left[ Y_{i,t}^{(s)} \mid \mathrm{do}\!\left(T_{i,t}^{(s)} = 1\right) \right]
 - \\\mathbb{E}\!\left[ Y_{i,t}^{(s)} \mid \mathrm{do}\!\left(T_{i,t}^{(s)} = 0\right) \right].
$}

The average treatment effect (ATE) aggregates these individual effects across all ego tweets and captures the average causal impact of prior exposure to stance $s$ on stance expression in the population:

\scalebox{0.9}{$
\mathrm{ATE}^{(s)}
= \mathbb{E}\!\left[ Y^{(s)} \mid \mathrm{do}\!\left(T^{(s)} = 1\right) \right]
 - \\
 \mathbb{E}\!\left[ Y^{(s)} \mid \mathrm{do}\!\left(T^{(s)} = 0\right) \right].
$}


ProEmb relies on a set of explicit assumptions: Rather than assuming that all confounders are directly measured, ProEmb leverages high-dimensional proxy variables constructed from observed ego and peer content that are informative of these latent factors. Thus, conditional on a suitable representation of these proxies, the unobserved confounders affecting both network formation and individual behavior are sufficiently accounted for, enabling identification of causal effects of exposure through the ego network~\cite{fatemi2025contagion}. 

\begin{figure*}[h]
\centering
\includegraphics[width=0.9\textwidth]{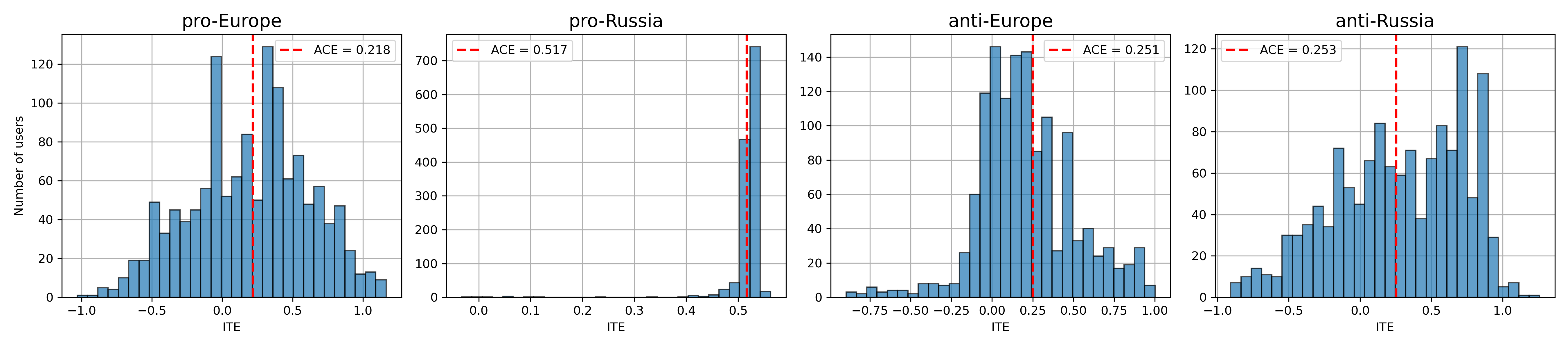} 
\caption{Individual treatment effect (ITE) of exposure to stance-aligned content on expression of similar stances in subsequent tweets, obtained with ProEmb.}
\label{fig:ite_distributions}
\end{figure*}

To construct the high-dimensional proxies used in ProEmb, we start from the full stance-annotated dataset and restrict it to X posts, due to the less consistent information across the other platforms. For each tweet, we retain the author ID, timestamp, text, stance label for the target category, and emotion scores (already available in the provided dataset). We then build an ego–peer structure from the retweet graph: for each retweet, we record the retweeter (ego) and the original author (peer), and group tweets chronologically by user. For every ego tweet, we gather (i) the ego’s own text and emotion scores at that time, and (ii) the collection of all past tweets by their peers (those authored before the ego tweet). Peer texts are concatenated and embedded, and peer emotion scores are averaged across tweets. This yields two feature vectors: an \emph{ego} feature vector (embedding of the current tweet plus its emotions) and a \emph{peer} feature vector (embedding of accumulated peer content plus mean peer emotions). Given the presence of multiple languages in the dataset, we use the multilingual model \texttt{paraphrase-multilingual-MiniLM-L12-v2}~\cite{reimers-2019-sentence-bert} to generate sentence embeddings.
In our implementation, we concatenate ego and peer features to form the proxy input, which is then encoded into a latent representation. This latent code acts as a balancing score: treated and control observations with similar latent representations are comparable under the proximal causal framework. 
Using these representations, ProEmb estimates causal effects by explicitly modeling counterfactual outcomes. For each target stance $s$, two outcome models are trained: one using ego tweets observed under exposure to that stance ($T_{i,t}^{(s)}=1$) and one using ego tweets observed without such exposure ($(T_{i,t}^{(s)}=0$). Each model predicts the probability that an ego tweet expresses the target stance. Counterfactual outcomes are obtained by applying these models crosswise: the model trained on exposed tweets is applied to unexposed tweets to predict outcomes as if they had been exposed, and vice versa. The ITE for a tweet is computed as the difference between these two predicted outcomes, holding observed covariates fixed. The ATE is obtained by aggregating ITEs across tweets, and is also estimated via inverse probability weighting in the latent space~\cite{fatemi2025contagion}.



\begin{table}[ht]
    \centering
    \caption{Average treatment effect (ATE) ProEmb estimates of exposure to stance-aligned content on subsequent stance expression. $n_1/n_0$ denotes the treated/control sample sizes.
    }
    \begin{tabular}{lcc}
        \toprule
        Stance & ATE (ProEmb) & $n_1/n_0$ \\
        \midrule
        Pro-Europe  & $0.218\pm0.034$ & $614/703$ \\
        Pro-Russia  & $0.517\pm0.0002$ & $21/1{,}296$ \\
        Anti-Europe & $0.251\pm0.043$ & $160/1{,}157$ \\
        Anti-Russia & $0.253\pm0.021$ & $840/477$ \\
        \bottomrule
    \end{tabular}
    \label{tab:ate_values}
\end{table}

Table~\ref{tab:ate_values} reports the ProEmb estimates averaged across
ten random seeds, together with the treated and control sample sizes. Across all stance categories, ProEmb yields positive and statistically significant average causal effect estimates, indicating that prior exposure to stance-aligned content increases the probability that a user subsequently expresses the same stance. The estimated ATEs are relatively consistent in magnitude: exposure to ``pro-Europe" content is associated with an average increase of $0.218$ in the probability of stance expression, while ``anti-Europe" and ``anti-Russia" exposures yield comparable effects of $0.251$ and $0.253$, respectively. The estimated effect for ``pro-Russia" exposure is substantially larger in magnitude ($0.517$), exceeding the effects observed for all other stances. However, this estimate should be interpreted with caution due to the relatively small number of treated observations in this category, which limits statistical precision and increases sensitivity to modeling assumptions.

Fig.~\ref{fig:ite_distributions} further illustrates the heterogeneity at the user level through the distribution of estimated ITE. Although the majority exhibit modest treatment effects, the distributions are right-skewed, revealing a subset of users who, when exposed, have an increased likelihood of stance-aligned posting. This pattern suggests that peer exposure does not uniformly influence behavior, but instead disproportionately affects a smaller group of highly responsive users. 
The list of hyperparameters and models used for ProEmb can be found in the Appendix.



%% file: sections/7-discussion.tex
\section{Discussion}
\label{sec:discussion}

Across platforms, the conversation is dominated by narratives related to European integration and opposition to Russian influence, particularly during periods of increased political interest around the election. This dominance is visible not only in aggregate volume but also in the structure of interaction, with these narratives occupying the most central positions in both retweet-based amplification and semantic alignment networks.  The Moldovan election was organized around a geopolitical discussion in which EU integration and Russian influence were central political issues. These asymmetries instead suggest that the online discourse we observe was not a balanced contest between equally visible parties. Instead, the conversation was structured around a dominant pro-European and anti-Russian frame, with alternative narratives appearing less visible and less structurally amplified.

Amplification patterns further differentiate how narratives propagate. Retweet networks exhibit a dense core composed of highly active accounts repeatedly amplifying a narrow set of messages, while semantic similarity networks reveal more selective alignment through near-duplicate framing within an event-scale temporal window. The observed structural differences imply that not all narratives benefit equally from synchronized amplification, even when coordinated-looking behavior is present. In particular, ``pro-Europe" and ``anti-Russia" narratives appear to benefit from both direct amplification of prominent political and media accounts and repeated semantic framing against Russian influence. At the same time, because the 1-hour window usually captures organic attention to trending topics~\cite{de2024detecting}, these patterns should not be interpreted as evidence of inauthentic coordination.

Beyond descriptive patterns, causal analysis provides evidence that exposure to stance-aligned content increases the likelihood of subsequent stance expression. The ProEmb framework identifies consistent positive effects for several stance categories. Prior work has shown that political discourse on social platforms is highly polarized, with users embedded in ideologically homogeneous communities that resemble echo chambers~\cite{conover2011political, garimella2018political, cinelli2021echo, cinus2022effect}. These studies document strong assortative mixing and selective exposure in political networks, but largely stop short of establishing whether exposure itself causally shapes political expression. Our results complement these studies by showing that, under the proximal identification assumptions and using observable exposure through retweet-based peers, exposure to stance-aligned content has a positive estimated effect on subsequent stance expression. This suggests that echo chambers are not only a structural property of political networks but also a potential reinforcement mechanism through which already visible narratives become repeatedly expressed and amplified over time.

Several limitations should be acknowledged. All causal estimates rely on observational data and proxy-based adjustment, which cannot fully eliminate bias from unmeasured confounders. Moreover, our treatment definition captures observable exposure through retweet-based peer ties, not all content that users may have encountered through algorithmic feeds, search, screenshots, external media, or offline communication. An additional source of risk lies in model misspecification. Estimating peer influence and contagion effects depends on the ability of the chosen model to capture the relevant exposure mechanisms, temporal dependencies, and latent confounding structure. Misspecified models can yield biased or misleading effect estimates, including overestimating or underestimating the contagion effects. These limitations call for cautious interpretation, particularly in politically sensitive contexts.

Overall, this work contributes empirical evidence on the interaction between political narratives, coordination mechanisms, and peer influence in a geopolitically salient election. By combining language-based annotation with network and causal methods, the study demonstrates the value of integrated analytical frameworks for understanding political communication in social media public environments.

%% file: sections/appendix.tex
\setcounter{section}{0}
\section{Broader Perspective and Ethics}
All data used in this study were publicly available and collected in accordance with the platforms’ terms of service. 
To minimize risks to users, all identifiable information was removed, and analyses were conducted on aggregated data only. Data were publicly collected in the United States and did not require user consent, leading us to assess the potential negative impacts of data use as minimal.
This analysis aims to provide positive societal value by improving the understanding of information campaign tactics. The findings may help researchers identify and mitigate similar forms of coordinated activity in future political contexts.
The presence of measurable stance contagion also raises important ethical considerations regarding the societal implications of social influence in online political environments. By demonstrating that exposure to stance-aligned content can causally affect subsequent expression, our findings highlight a mechanism through which political narratives may reinforce themselves over time. In principle, such dynamics could be exploited in manipulative or strategic ways, for example, by deliberately amplifying targeted messages to influence opinion formation, mobilization, or polarization.
It is important to emphasize that we do not propose, develop, or deploy tools for political persuasion, targeting, or manipulation. Rather, our objective is to document and quantify the existence of contagion effects in real-world electoral discourse, contributing empirical evidence to ongoing discussions about the role of social influence in digital political communication. Although focused on the 2024 Moldovan presidential election, we can expect that the insights from this study can be generalized to other political events.

\section{Data Extraction and Query Strategy}
\label{app:data_extraction}

Election-related social media data were collected via a commercial third-party social media monitoring provider that aggregates publicly available content from multiple platforms under platform Terms of Service and applicable data protection regulations, including GDPR. The provider supports Boolean keyword queries, metadata-based filtering, and engagement-based retrieval, while enforcing platform-specific limits on data volume, retention, and redistribution. The scripts with relative keywords are available in the repository.



\noindent\textbf{Query Design.} To capture a broad range of election-related discourse, a set of complementary query templates was employed, targeting (i) political actors and parties, (ii) election- and referendum-related keywords, (iii) multilingual variants of core election terms, and (iv) interaction-based content such as replies, retweets, and quoted posts involving prominent Moldovan political accounts. Content unrelated to politics (e.g., sports or entertainment) was explicitly excluded where possible. In addition, some queries targeted engagement with specific high-profile accounts to capture secondary discussion and amplification dynamics beyond original posts. Table~\ref{tab:query_types} summarizes the main categories of queries used during data collection. \\
\noindent\textbf{Scope and Limitations.} The extraction strategy was designed to capture as much relevant discourse as possible within practical and legal constraints. However, platform-specific limitations, uneven data availability, and throttling policies may have resulted in different coverage across users. Furthermore, because several queries directly name candidates, parties, Russia/EU-related actors, and engagement with selected accounts, the retrieval design may affect both stance prevalence and the apparent centrality of narratives. The authors were not given the provider's identity, collection budget, complete retrieval logic, or platform-specific rate limits. As a result, findings should be interpreted as reflecting observed discourse within the collected dataset rather than population-level estimates of all online political activity during the election period.

\begin{table}[ht]
    \centering
    \caption{Summary of query categories used for election-related data extraction.}
    \small
    \begin{tabular}{p{1.6cm} p{6cm}}
        \toprule
        Query category & Description \\
        \midrule
        Actor-based & Mentions of prominent Moldovan political figures, candidates, and party leaders, including multilingual name variants. \\
        Party-based & References to major political parties and coalitions participating in the election, expressed in multiple languages. \\
        Election keywords & Proximity-based queries capturing elections, voting, referendum-related terms, and electoral procedures across languages. \\
        Interaction-based & Posts engaging with or authored by key political accounts (e.g., replies, retweets, quotes), used to capture amplification and discussion dynamics. \\
        Location-based & Content associated with Moldova or major cities (e.g., Chișinău), combined with political keywords to increase contextual relevance. \\
        Platform-specific & Queries tailored to platform constraints (e.g., site-restricted searches or first-post filters) to improve coverage consistency. \\
        \bottomrule
    \end{tabular}
    \label{tab:query_types}
\end{table}

\begin{figure*}[h]
\centering
\includegraphics[width=\textwidth]{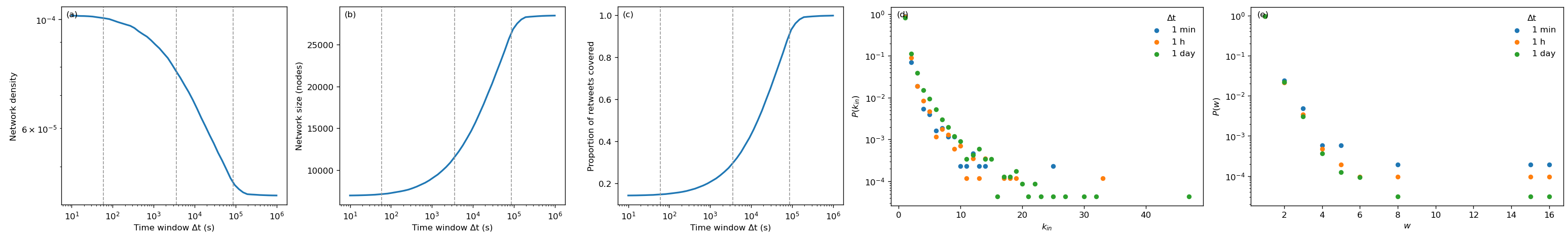}
\caption{Sensitivity of the retweet-based synchronized amplification network to the temporal threshold $\Delta t$. Panels (a)--(c) show how network density, network size, and retweet coverage change as the threshold increases. Dashed vertical lines mark the representative thresholds of 60\,s, 1\,hour, and 24\,hours.}
\label{fig:retweet_threshold_sensitivity}
\end{figure*}


\section{Political Stance Classification and LLM Prompting}
Before consensus adjudication, the two annotators agreed on 84.6\% of
the individual label decisions. Mean label-specific Cohen's $\kappa$
was 0.324. All disagreements were subsequently reviewed and resolved
through discussion to establish the final reference annotations. For classification, we explored multiple prompting strategies commonly used for this task to assess robustness and sensitivity to prompt design. Specifically, we evaluated zero-shot prompting, in which the model is asked to infer stance labels solely from natural-language definitions; few-shot prompting, which supplements these definitions with a small number of labeled examples; Chain-of-Thought (CoT) prompting, which encourages the model to reason explicitly before producing a label~\cite{sahoo2024systematic}; and a combined CoT + few-shot approach that integrates labeled demonstrations with step-by-step reasoning instructions.
The final prompting setup used in our analysis frames stance classification as a structured instruction-following task. For each tweet, the model is required to output a fixed-format JSON array indicating whether the content has a ``pro", ``anti", or ``neutral" stance in the political context of the 2024 Moldovan presidential election, along with a truncated tweet snippet, the tweet ID, and an index (for subsequent merging of the stances in the database). The prompt enforces a two-step decision process—first determining whether Russia (and/or the presidential candidate Stoianoglo) or Europe (and/or the presidential candidate Sandu) is mentioned in a relevant political context, and then assigning sentiment labels conditionally on that mention, while allowing for mixed sentiment when explicitly present. To ensure consistency and machine-readability, the prompt strictly constrains output formatting and provides illustrative examples covering positive, negative, neutral, mixed, and non-relevant cases. Furthermore, for each axis, the classifier returns four binary indicators: pro, anti, neutral, and non-relevant. Pro and anti may co-occur when a post contains mixed framing, and neutral is used only when the target is politically relevant but neither positive nor negative framing is present. Using the consensus-adjudicated annotations
as the reference, the LLM classifier achieved pooled precision, recall, and
F1 scores of 0.810. Performance was particularly strong for identifying
content unrelated to Russia and Europe (F1${}=0.932$ and 0.885,
respectively), while the anti-Europe and anti-Russia categories obtained
F1 scores of 0.750 and 0.700. The macro-F1 across all eight labels was
0.620, and variation across categories partly reflects the limited number
of positive examples available for several comparatively rare stances. \\
For readability, the complete prompt text and implementation details are available in the repository.

\section{Additional Retweet-Network Analyses}

Fig.~\ref{fig:retweet_threshold_sensitivity} reports the sensitivity of the retweet-based synchronized amplification network to the temporal window $\Delta t$~\cite{de2024detecting}. As expected, increasing $\Delta t$ expands the network and increases the fraction of retweets covered, while decreasing network density. The 60,s threshold provides the strictest synchronization signal, but covers only a limited share of retweet activity and yields a comparatively small network. In contrast, the 24,hour threshold approaches saturation, covering nearly all retweets, which makes it more likely to capture general trending-topic dynamics rather than temporally localized amplification.
The 1,hour threshold lies between these two extremes. It substantially increases retweet coverage and network size relative to the 60,s window, while remaining below near-saturation, observed at 24,hours. The in-degree and edge-weight distributions remain heavy-tailed across the three representative thresholds, suggesting that the amplification structure is not driven solely by a few isolated edges.

To contextualize the structure of the rapid-retweet network,
Table~\ref{tab:central_accounts} lists selected high-degree accounts
from its largest connected component. The examples include political
figures, news organizations, journalists, activist accounts, and
satirical accounts. The account descriptions are based on publicly available profile information and observed content.

\begin{table}[h]
    \centering
    \caption{Selected high-degree accounts in the rapid retweet network. Descriptions are based on publicly available account information and observed content.
}
    \begin{small}
    \begin{tabular}{p{3.2cm} p{4.2cm}}
        \toprule
        Account & Description \\
        \midrule
        @SanduMaiaMD & Official account of President Maia Sandu \\
        @vonderleyen & Official account of Ursula von der Leyen, President of the European Commission \\
        @JuliaDavisNews & Journalist and columnist monitoring Russian state media \\
        @wartranslated & Account translating and reposting Russo-Ukrainian war footage \\
        @nexta\_tv & Eastern European opposition-aligned news outlet \\
        @IgorSushko & Ukrainian--American political commentator on Eastern Europe \\
        @IamDenya\_de & President of IPN Press Agency, Moldovan news outlet \\
        @DarthPutinKGB & Satirical account critical of Vladimir Putin \\
        @YourAnonCentral & Anonymous political activists \\
        @visegrad24 & International news account focused on Central-Eastern Europe \\
        \bottomrule
    \end{tabular}
    \end{small}
    \label{tab:central_accounts}
\end{table}

\section{Training Details and Computational Resources}

All experiments were conducted on an internal laboratory server. The system is equipped with a single AMD EPYC 7662 processor (64 cores, 128 threads) and 256\,GB of system memory. GPU-accelerated workloads were run on an NVIDIA RTX A5000 GPU with 24\,GB of VRAM, using CUDA~12.8. 

For topic modeling, we used BERTopic~\cite{grootendorst-arxiv22}. Sentence embeddings are generated using a transformer-based multilingual embedding model, followed by dimensionality reduction with UMAP and clustering within the BERTopic framework. The UMAP configuration uses \texttt{n\_neighbors=15}, \texttt{n\_components=5}, \texttt{min\_dist=0.0}, cosine distance, and a fixed random seed for reproducibility. BERTopic is instantiated with an n-gram range of $(1,3)$, a minimum topic size of 15 documents, and a target of 40 topics. These hyperparameters were chosen to balance topic coherence and granularity and were not optimized against downstream tasks.
For causal analyses, we used sentence embeddings generated with the \texttt{paraphrase-multilingual-MiniLM} \texttt{-L12-v2} model~\cite{reimers-2019-sentence-bert}. This choice enables a shared semantic representation across languages, which is essential given the multilingual nature of the dataset. We followed the ProEmb modeling choices and hyperparameter settings recommended in the original ProEmb paper, without additional tuning~\cite{fatemi2025contagion}. However, we use gradient-boosted decision trees implemented with LightGBM as the outcome model and rely on text embeddings as covariates instead of bag-of-words representations, which are less suitable for multilingual data. To quantify sensitivity to stochastic model fitting, we repeated ProEmb across ten random seeds and report the mean and between-seed standard deviation.

\section{Topic Prevalence and Macro-Topic Mapping}
\begin{figure}[h]
    \centering
    \includegraphics[width=\columnwidth]
    {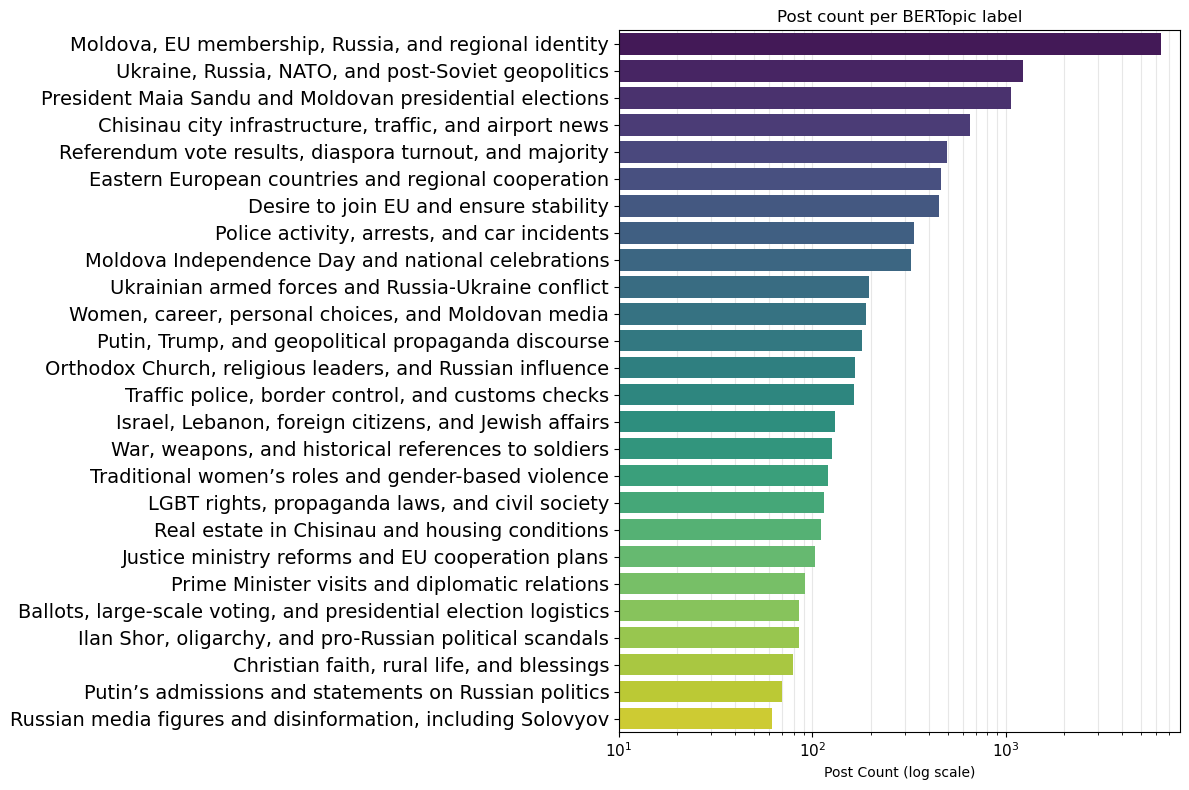}
    \caption{Topic prevalence across election-related posts, ordered by frequency and shown on a logarithmic scale. The distribution exhibits a pronounced long tail, with a small number of dominant topics and many low-frequency topics.}
    \label{fig:topics}
\end{figure}

Figure~\ref{fig:topics} presents the complete ranked distribution of
the retained election-relevant topics. Topic prevalence is strongly
concentrated: the three most frequent topics account for 64\% of the
observed posts, while the eight most frequent topics account for 82\%.
The remaining topics form a long tail of comparatively infrequent
themes. This concentration motivates both the main-text focus on topics
representing at least 2\% of posts on one platform and the higher-level
macro-topic aggregation reported below.

To facilitate higher-level analysis and cross-platform comparison, we manually aggregate generated topics into four macro-topics. This aggregation is based on a qualitative inspection of each topic, aiming to group semantically coherent themes while preserving interpretability. Table~\ref{tab:topic_macro_mapping} reports the mapping between topics and the corresponding macro-topic. Considering the qualitative nature of the aggregation process, language-by-macro-topic comparisons should be interpreted as exploratory summaries and not as validated category measurements.

\begin{table}[ht]
\centering
\caption{Manual aggregation of BERTopic topics into macro-topics based on qualitative inspection of topic content.}
\footnotesize
\begin{tabular}{p{1.6cm} p{6cm}}
\toprule
\textbf{Macro-Topic} & \textbf{Mapped Topic Descriptions} \\
\midrule
\multirow{8}{*}{\parbox{1.6cm}{\centering Geopolitics \& EU Integration}}
& Ukraine, Russia, NATO, and post-Soviet geopolitics \\
& Eastern European countries and regional cooperation \\
& Referendum results and diaspora turnout \\
& EU accession and political stability \\
& Justice reforms and EU cooperation plans \\
& Putin, Trump, and geopolitical propaganda discourse \\
& Statements on Russian politics \\
& Diplomatic visits and international relations \\
\midrule
\multirow{5}{*}{\parbox{1.6cm}{\centering Domestic Politics \& Governance}}
& Moldova, EU membership, Russia, and regional identity \\
& President Maia Sandu and presidential elections \\
& Russian media figures and disinformation narratives \\
& Ilan Shor, oligarchy, and pro-Russian political scandals \\
& Ballots, large-scale voting, and election logistics \\
\midrule
\multirow{7}{*}{\parbox{1.6cm}{\centering Society \& Civic Life}}
& Chisinau infrastructure, traffic, and airport news \\
& Independence Day and national celebrations \\
& Police activity, arrests, and public safety \\
& Border control and customs checks \\
& Gender roles and gender-based violence \\
& LGBT rights and civil society debates \\
& Housing conditions and real estate in Chisinau \\
\midrule
\multirow{6}{*}{\parbox{1.6cm}{\centering Culture, Religion, \& Conflict}}
& Orthodox Church and religious leadership \\
& Russia--Ukraine war and armed forces \\
& War narratives and historical references \\
& Christian faith, rural life, and traditions \\
& Women, careers, and media narratives \\
& Israel, Lebanon, and foreign affairs \\
\bottomrule
\end{tabular}

\label{tab:topic_macro_mapping}
\end{table}